\documentclass[aps,prl,amsmath,amssymb,reprint,superscriptaddress,showpacs]{revtex4-2}      
        
\usepackage{graphicx}   
\usepackage{dcolumn}
\usepackage{bbold}
\usepackage[T1]{fontenc}
\usepackage[utf8]{inputenc}
\usepackage{lmodern}
\usepackage{color}
\usepackage[dvipsnames]{xcolor}
\usepackage[colorlinks=true, linkcolor=blue, citecolor=blue, urlcolor=blue,breaklinks=true]{hyperref}

\begin{document}

\title{Certified Random Number Generation from Quantum Steering}

\author{Dominick J. Joch}
\affiliation{Centre for Quantum Dynamics and Centre for Quantum Computation and Communication Technology, Griffith University, Brisbane, Queensland 4111, Australia}

\author{Sergei Slussarenko}\affiliation{Centre for Quantum Dynamics and Centre for Quantum Computation and Communication Technology, Griffith University, Brisbane, Queensland 4111, Australia}

\author{Yuanlong Wang}\affiliation{Centre for Quantum Dynamics and Centre for Quantum Computation and Communication Technology, Griffith University, Brisbane, Queensland 4111, Australia}

\author{Alex Pepper} \affiliation{Centre for Quantum Dynamics and Centre for Quantum Computation and Communication Technology, Griffith University, Brisbane, Queensland 4111, Australia}

\author{Shouyi Xie}\affiliation{Centre for Quantum Computation and Communication Technology, School of Physics, The University of New South Wales, Sydney, NSW 2052, Australia}
\affiliation{Current address: School of Physics, The University of Sydney, Camperdown, NSW 2006, Australia}

\author{Bin--Bin Xu}\affiliation{Centre for Quantum Computation and Communication Technology, School of Physics, The University of New South Wales, Sydney, NSW 2052, Australia}
\affiliation{ Current address: Beijing Key Laboratory for Precision Optoelectronic Measurement Instrument and Technology, School of Optics and Photonics,
Beijing Institute of Technology, Beijing 100081, China}

\author{Ian R. Berkman}\affiliation{Centre for Quantum Computation and Communication Technology, School of Physics, The University of New South Wales, Sydney, NSW 2052, Australia}

\author{Sven Rogge}\affiliation{Centre for Quantum Computation and Communication Technology, School of Physics, The University of New South Wales, Sydney, NSW 2052, Australia}

\author{Geoff J. Pryde}
\email[]{g.pryde@griffith.edu.au}
\affiliation{Centre for Quantum Dynamics and Centre for Quantum Computation and Communication Technology, Griffith University, Brisbane, Queensland 4111, Australia}

\date{\today}

\begin{abstract}

The ultimate random number generators are those certified to be unpredictable---including to an adversary. The use of simple quantum processes promises to provide numbers that no physical observer could predict but, in practice, unwanted noise and imperfect devices can compromise fundamental randomness and protocol security. Certified randomness protocols have been developed which remove the need for trust in devices by taking advantage of nonlocality. Here, we use a photonic platform to implement our protocol, which operates in the quantum steering scenario where one can certify randomness in a one-sided  device independent framework. We demonstrate an approach for a steering-based generator of public or private randomness, and the first generation of certified random bits, with the detection loophole closed, in the steering scenario.\\

\end{abstract}

\maketitle

Randomness is an essential resource in many applications from simulation to cryptography. For applications where one cares about security, \textit{certified} private randomness is required---randomness that is guaranteed to be not predictable to an adversary or physical observer~\cite{Acin2016, superluminal}. Purportedly-random numbers can be tested for uniformity and the presence of predictable patterns,  but such tests can be satisfied by some pseudo-random number generators (Pseudo-RNG)~\cite{Acin2016, instrumental2020}.  As these statistical tests can be passed by sets of numbers of deterministic origin, one cannot rely on them to assert unpredictability. Instead one must certify randomness in the generation process itself~\cite{qrng2017}. 

Quantum phenomena display intrinsic randomness and can thus serve as quantum random number generators (QRNG). Standard QRNG~\cite{anu, Gehring2021, Nie2015, Haylock2019} operates in a trusted-device scenario that relies on assumptions about, and accurate modelling of,  physical devices. Imperfections are susceptible to exploitation by adversaries~\cite{Acin2016, superluminal, Pironio2010,Liu2021,avesani2021} as they can carry side information. Classical side information comes from sources like thermal and electronic noise, which may be of a malicious nature (known to, or controlled by, an adversary), and quantum side information arises from correlation with another quantum system. Hardware failures in devices could also compromise the output~\cite{qrng2017} and noise will inevitably be introduced by experimental imperfections~\cite{instrumental2020}.

Certified random numbers are produced from a process that any physical observer cannot perfectly predict, under a minimal set of assumptions---the fewer and weaker the assumptions, the stronger the security. Certification and post-processing are necessary to acquire private randomness that is independent of side information~\cite{qrng2017,marangon2017}.  Side information can be accounted for within the strategies of an adversary, hence randomness is certified by finding an upper bound on an adversary’s predictive ability.

To minimize device related assumptions as much as experimentally possible,  randomness can be certified by device independent QRNG (DIQRNG) protocols in a Bell test (or instrumental~\cite{Chaves2018, instrumental2020}) scenario~\cite{Acin2016, Pironio2010, qrng2017}, offering the highest possible security when loopholes are closed~\cite{avesani2021}. The realization of such protocols has been achieved only recently~\cite{Pironio2010, superluminal, Liu2018, Shen2018}, with extreme security following the first strong-loophole-free Bell tests~\cite{Hensen2015, Shalm2015,significantloopholefree}. Currently, work is still progressing towards reaching the rates desired for commercial applications with loopholes closed~\cite{ Zhang2020, Shalm2021,Liu2021}. DIQRNG is technically demanding as very high detection efficiencies and low noise are necessary to certify randomness with closed loopholes, which poses a challenge to achieving high rates. Therefore, many semi-device-independent (SDI) protocols, exploiting the trade-off between security and ease of implementation, have been developed~\cite{qrng2017,avesani2021, marangon2017, passaro,  Zhang2021}.\\

 Here we demonstrate an experimental implementation of a QRNG protocol, based on the steering nonlocality task~\cite{Wiseman2007, steeringreview}, with the detection loophole closed, and extract certified random numbers with a high security against general attacks of a quantum adversary. Steering nonlocality can be used to realize one-sided device independent (1SDI) QRNG~\cite{passaro, Mattar2017}, that offers improved security over trusted-device QRNGs by removing many device related assumptions.  Due to steering having trust assumptions different from a Bell inequality, states needed for DIQRNG are steerable, but not all steerable states violate a Bell inequality.  It provides greater noise tolerance and is more robust to loss, allowing randomness to be certified at lower efficiencies than DIQRNG~\cite{passaro}. We show the first steering-based QRNG generation of random bits in such a regime.

\section{Results}

\begin{figure}
\includegraphics[width=8cm]{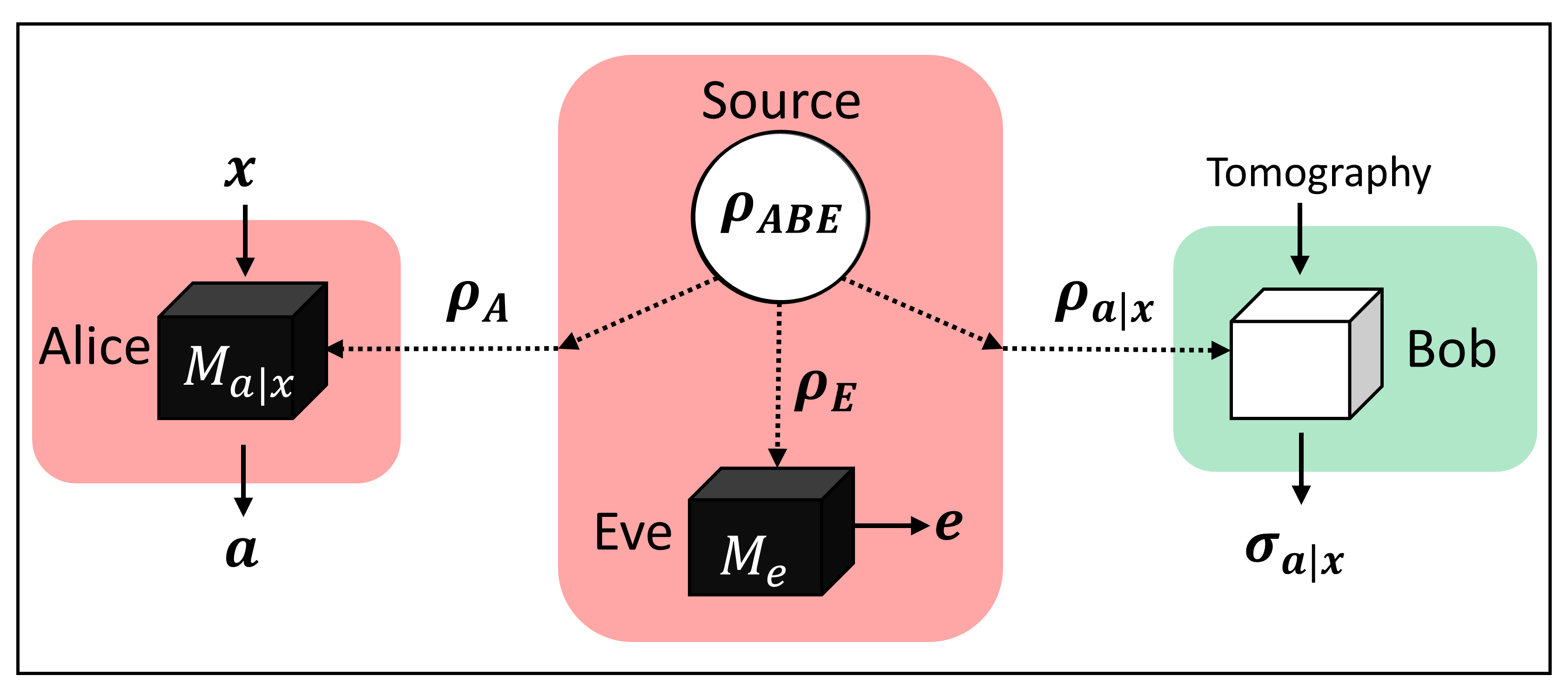}
\caption{\textbf{Adversarial scenario for randomness certification.} The eavesdropper, Eve, is assumed to control the source. It is assumed that, while in principle the source distributes bipartite states to Alice and Bob, she may in fact distribute a tripartite state such that $\rho_{\rm{AB}}= \rm{Tr}_{\rm{E}} \rho_{\rm{ABE}}$. She may perform a measurement $M_e$ on her subsystem, acquiring the outcome $e$ which is her guess for Alice's outcome, and both $e$ and $M_e$ are unknown to Bob. Alice selects a setting $x$ and acquires outcome $a$ from the measurement $M_{a|x}$. Bob performs quantum state tomography and obtains an assemblage $\sigma_{a|x}$ of conditional states.}
\label{eve}
\end{figure}

\textbf{1SDI Randomness Certification.} First, we introduce the quantum steering scenario. Consider two parties named Alice and Bob who receive a bipartite state $\rho_{\rm{AB}}$ from some untrusted source. One-sided device independence comes from one party (Bob, say) being trusted and the other, Alice,  being untrusted. Her measurement device is treated as a black box with classical inputs $x \in \{1, ...,\mathcal{M}\}$ for $\mathcal{M}$ measurement settings, and classical outcomes $a \in \{0, 1, \text{Ø}\}$  where Ø is the null outcome to account for experimental losses. A trusted Bob is not malicious and has full knowledge of the inner workings of his measurement device. Bob accepts quantum mechanics to be valid and can perform quantum state tomography to construct an \textit{assemblage}---a set of unnormalized quantum states conditional on Alice's settings and outcomes~\cite{steeringreview}. Properties of the assemblage determine whether quantum steering pertains. Specifically, in the assemblage picture, quantum steering can be tested via a semidefinite program (SDP)~\cite{sdpsteeringrev,steeringreview} (see Methods), an approach that proves useful in randomness certification. It is also possible to construct steering protocols where the need to trust Bob is greatly reduced by using quantum instructions~\cite{Kocsis2015}, although trust in quantum mechanics is still required.

To certify the local randomness of Alice's outcomes, an adversarial situation is considered where in a given trial, an eavesdropper (Eve) attempts to predict Alice’s outcome (Fig.~\ref{eve}). The trials are assumed to be independent and identically distributed with respect to Eve’s strategy~\cite{passaro}. 
 If Eve's guessing probability $P_{g}(x^*)$  is less than unity she cannot perfectly predict the outcome of Alice, and some randomness is certified as quantified by the min-entropy
\begin{equation}
 H_{\rm{min}}=-\log_{2}[P_{g}(x^*)].
\end{equation}

The upper bound on the certified randomness is found by optimizing Eve's guessing strategy to maximize $P_{g}(x^*)$. The source of bipartite states, being untrusted, may be in Eve's possession so we assume the states $\rho_{\rm{AB}}$ are correlated with another quantum system held by Eve, as in  ~Fig.~\ref{eve}. Since Eve's outcome is unknown, Bob's observed assemblage theoretically is of the form 
\begin{equation}
\sigma_{a|x} = \sum_{e} \sigma^{e}_{a|x}=\sum_{e} \rm{Tr}_{\rm{AE}} \mathnormal{[(M_{a|x} \otimes \mathbb{1}_{B} \otimes M_{e})}\rho_{\rm{ABE}}].
\end{equation}
 Eve's strategy is accounted for in Bob's assemblage and so the optimisation is done with respect to $\{\sigma^e_{a|x}\}$---rather than $\rho_{\rm{ABE}}$, $\{M_{a|x}\}$ and $\{M_{e}\}$---by solving a semidefinite program (SDP)~\cite{passaro,sdpsteeringrev}:
\begin{equation}
\label{SDP1}
\vspace{-0.4cm} \underset{ \{\sigma^{e}_{a|x}\} _{a,x,e}}{\text{max}} \hspace{0.5cm}  P_g(x^{*})=\sum_{e} \rm{Tr}\mathnormal{(\sigma^{e}_{a=e|x^{*}})}, 
\end{equation}
$$\vspace{-0.4cm} \text{s.t.}  \hspace{1cm}\sum_{e}\sigma^{e}_{a|x}=\sigma_{a|x}   \hspace{1cm}   \forall{~a,x},$$
$$\vspace{-0.4cm}\sum_{a}\sigma^{e}_{a|x}=\sum_{a}\sigma^{e}_{a|x'}  \hspace{1cm}    \forall{~e,x\neq x'},$$
$$\vspace{-0.05cm}\sigma^{e}_{a|x} \geq 0 \hspace{1cm}    \forall{~a,x,e}.$$
The first constraint ensures compatibility with Bob's assemblage. The second enforces the non-signaling condition, that is, to disallow measurement settings at one party to influence outcomes at another. The third is a positive semidefinite constraint to ensure $\{\sigma^{e}_{a|x}\}$ consists of valid quantum states~\cite{passaro}. \\

As in the Bell scenario (see Ref.~\cite{larsson2014} for loopholes in Bell tests), certain assumptions open loopholes which would allow for nonlocality to be falsely verified while permitting a local causal explanation~\cite{Wittmann2012}. The detection loophole appears under the fair sampling assumption that the statistics of the detected sample accurately represents the total sample. However, loss in the untrusted device may constitute a cheating strategy, therefore a certain heralding efficiency (the probability of one party detecting given that the other party detects) is demanded of Alice to close this loophole. There is no such requirement upon Bob, unlike the Bell scenario.   \\

\textbf{QRNG Protocol.}
A certified random number generator is a two-stage protocol: entropy accumulation followed by randomness extraction. Our protocol involves the parties Eve (source), Alice, Bob, and another (necessarily trusted) party---sometimes called Victor the verifier---being the user who has access to the data recording and processing devices. In the first stage we acquire two sets of experimental data---that needed for the certification step, and another set which provides many weakly random bits. To do this, many trials of the protocol are performed. Each is structured as in Fig.~\ref{eve}, where: a source distributes a state; Alice  and Bob perform measurements;  and the outcomes are recorded. From the experimental certification data, an assemblage is determined and used with the
SDP of Eq.~(\ref{SDP1}) to obtain the min-entropy, $H_{\text{min}}$, that quantifies the randomness certified in the weakly random string. The protocol passes (i.e. is able to generate certified randomness) if $H_{\text{min}}$  is non-zero and the number of extractable bits is $m\geq1$. Conditional on passing, we or Victor (the user) apply a randomness extractor which is an algorithm that produces random numbers from the raw string. These random bits are certified, which means that an adversary cannot predict them provided they are kept private after generation and the other assumptions of our protocol hold. Our QRNG operates in a one-sided device independent framework where minimal assumptions are made concerning the physical nature of untrusted devices, measurements and states. \\

\begin{figure*}
\includegraphics[width=18cm]{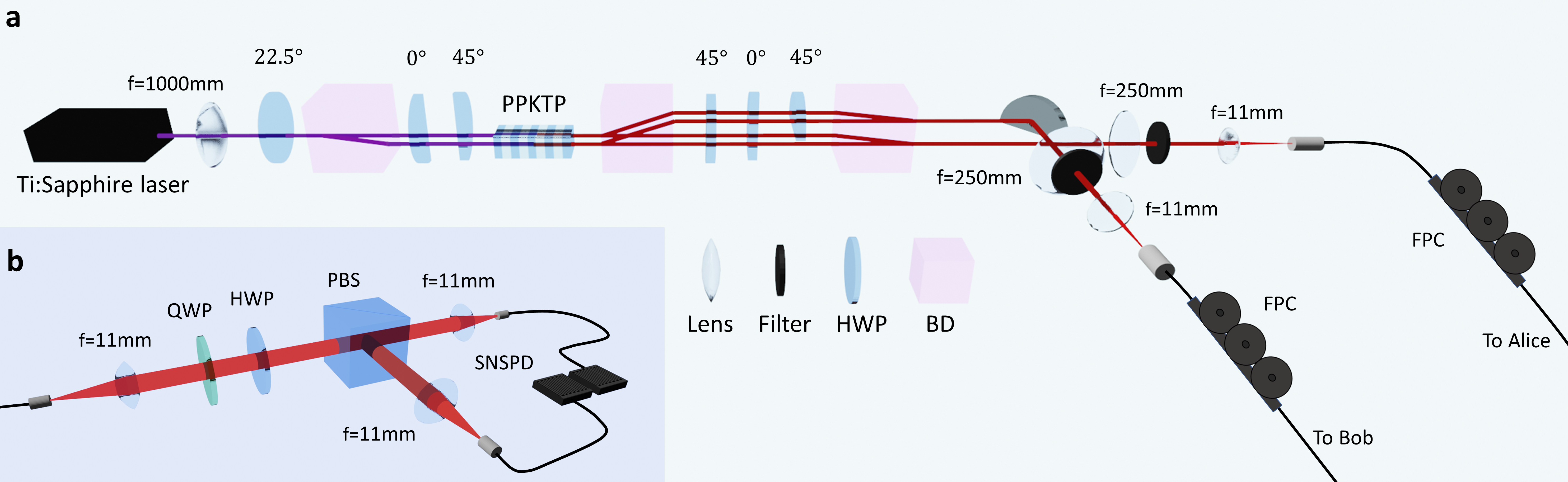}
\caption{\textbf{Experimental Setup. } \textbf{a.} The source of bipartite states.  A continuous wave 775nm pump from a Ti:Sapphire laser is focused in two locations inside a periodically poled $\text{KTiOPO}_4$ (PPKTP) crystal, creating telecom photons in pairs by type-II spontaneous parametric down conversion (SPDC). The process occurs within a Mach-Zehnder interferometer comprised of beam displacers (BDs) and a series of cut half waveplates (HWPs) to produce entangled states $\frac{1}{\sqrt{2}}\left(\left.|00\right\rangle+ e^{-i\phi}\left.|11\right\rangle\right)$ by interfering the two SDPC events ~\cite{sourcepaper}. The angles of the HWPs relative to their optical axis are annotated. The photon pairs are coupled into single mode fiber (SMF) and sent to the measurement devices of the two parties. Fiber polarization controllers (FPCs) are used to correct unwanted in-fiber transformations of the qubits, and to implement local unitary rotations to prepare each of the four canonical Bell states. 
\textbf{b.} Alice's and Bob's measurement devices.  Both Alice and Bob have the same physical setup; the combination of a quarter waveplate (QWP), half waveplate (HWP) and polarizing beam splitter (PBS) is used to measure the polarization state of the photons in different settings. Both output modes of the PBS are coupled into SMF and finally detected by superconducting nanowire single photon detectors (SNSPDs)~\cite{Marsili13}. Detection events are recorded by trusted devices of the user (time taggers, classical computers).}
\label{setup}
\end{figure*}

\textbf{Experimental implementation}
The experimental realization of the protocol---which uses photonic polarization qubits---is shown in Fig.~\ref{setup}. Our source prepares bipartite entangled states of  telecommunications-wavelength photons created by spontaneous parametric downconversion (SPDC). A quantum state fidelity of $\mathcal{F} = 0.9933\pm0.0005$ with $\left.|\Psi^-\right\rangle=\left(\left.|01\right\rangle-\left.|10\right\rangle\right)/\sqrt2$ was recorded for one data set; other data sets corresponded to generated states with comparable singlet-state fidelities. The photon pairs are coupled into optical fiber and sent to the measurement devices of the two parties (Fig.~\ref{setup}b). 

\begin{figure*}
\includegraphics[width=15cm]{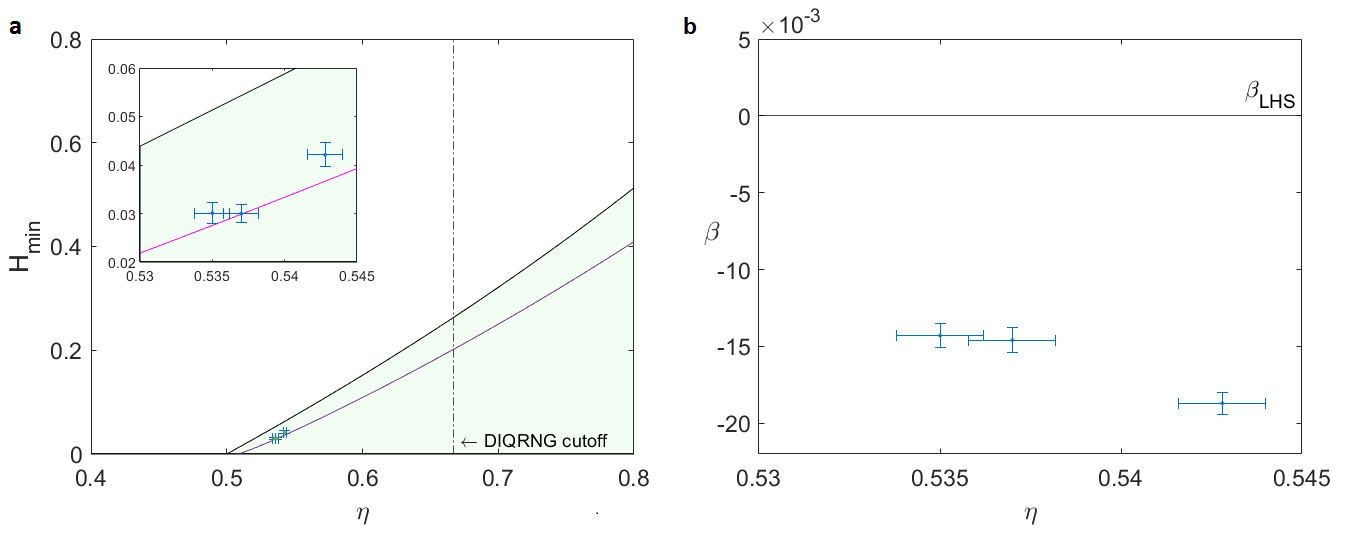}
\caption{\textbf{ Randomness Certification.} \textbf{a.} Certified min-entropy, $H_{\text{min}}$, against the heralding efficiency $\eta$ of Alice for multiple runs of the protocol.  The black line is the theoretical 1SDI bound for a maximally entangled two-qubit  Bell state, with a heralding efficiency threshold of 0.5. The purple line is for the Werner state $0.99\left.|\Psi^-\right.\rangle\left.\langle\Psi^-|\right. + 0.01 \mathbb{1}/4$. The dashed line is the lowest theoretical device independent threshold of $\frac{2}{3}$. 
\textbf{b.} Steering parameter $\beta$ versus heralding efficiency of Alice for the three data sets, showing a violation of the steering inequality Eq.~(\ref{ineq}), $\beta := \rm{Tr} \mathnormal{\sum_{a,x} F_{a|x}\sigma_{a|x} \geq} 0$.
}
\label{hmin}
\end{figure*}

In the implementation of the protocol, a certification data acquisition (later used to find $H_\textrm{min}$) is performed, followed by RNG data acquisition---these steps make up the accumulation stage. Bob's device allows him to measure in three complementary bases and use the data to perform quantum state tomography to determine his local states for each of the $\mathcal{M}=2$ settings (X, Z) and outcomes $a \in \{0,1,\text{Ø} \}$ of Alice. The closest physical assemblage for Bob's data is obtained with a maximum likelihood reconstruction to ensure the non-signaling condition (see Eq.~(\ref{SDP1}) conditions) is satisfied even in the presence of statistical noise arising from finite Poissonian data. The semidefinite program Eq.~(\ref{SDP1}) is solved to certify the amount of randomness present in the weakly random data. In Fig.~\ref{hmin}a, we compare our certification results with the theoretical bounds for 1SDI~\cite{passaro} and DI randomness certification methods. The corresponding steering inequality violations (see Methods) are shown in~Fig.~\ref{hmin}b. We obtained the highest min-entropy of $H_{\text{min}}=0.042\pm0.003$, at a heralding efficiency of $\eta_{\text{Alice}}=0.543\pm0.001$. Our results demonstrate randomness certification below the lowest heralding efficiency bounds of DI protocols at $2/3$~\cite{Shalm2015} , showing the advantage of the one-sided scheme to add experimental robustness. 

During the RNG data acquisition, detection is time tagged to record the exact time of a detection event and the detector channel. The raw time-tag data is post-processed; thanks to our trust in Bob, we can post-select the successful trials as coincidence events, where the parties detect a pair of photons within a $3~\rm{ns}$ window. We convert Alice’s outcomes into a string of bits by assigning detection channels to binary values, ignoring null outcomes (which does not compromise the security as long as the certification step, which takes into account null events, has been passed). 
We note that we do not need to assume that Alice's measurements are performed accurately or even at all, as long as her ``measurement'' strategy is the same during the certification and data acquisition phases. (In principle, Bob can tell her after the fact which trials contributed to certification and which to data acquisition, so no loophole need be opened.)

In the randomness extraction stage, weakly random data is post-processed by a classical algorithm to acquire certified random bits. We use an implementation of Trevisan’s extractor~\cite{instrumental2020, Trevisan2001}. This algorithm is a \textit{quantum-proof strong randomness extractor}, meaning that it is secure against both classical and quantum side information, and that the seed randomness is not consumed and can be reused~\cite{tmps2012}. Our extractor program is modified from the code of Ref.~\cite{instrumental2020}, which is based on the construction devised in Ref.~\cite{tmps2012}, and is discussed further in the Methods. In the case that the protocol passes the certification test, the extractor takes as input the uniform seed, min-entropy, error parameter $\epsilon$ and weakly random $(n,k)$-source (defined in the Methods), and outputs a string of certified random bits.  Note the extractor is independent of the general scheme of our protocol, so any suitable quantum-proof strong extractor may be used.

From our datasets we extracted certified random bits using seed bits obtained from a trusted-device QRNG beacon~\cite{anuweb, anu}.  From the largest dataset($H_{\text{min}}=0.030$, $\eta=0.535$) we could have, in principle, extracted 111,035 certified random bits uniform to within $2^{-64}$ with 8,126,464 seed bits.
 From the other two datasets in principle we could extract 7018 ($H_{\text{min}}=0.042$) and 8073 ($H_{\text{min}}=0.030$, $\eta=0.537$)  certified random bits uniform to within $2^{-64}$, with 4,456,448 and 4,718,592 seed bits respectively.
This extractor has low entropy loss versus error and performs well for low min-entropies. However, it requires a large seed, so the full computation was not performed for $\epsilon=2^{-64}$ . Using Trevisan's extractor with large datasets also becomes computationally demanding~\cite{Shen2018, Shalm2021, tmps2012} and impractical for low-latency RNG. A solution is to perform extraction with smaller sets of weakly random data to produce the output sequence in blocks with greatly reduced runtime and seed requirement due to the strong extractor property~\cite{post2013}. 
By processing 20 kb at a time we extracted 6489 certified random bits uniform to within $10^{-6}$ generated in 754-bit blocks, with 180,224 seed bits, from the $H_{\text{min}}=0.042$ dataset, and 7131 certified random bits uniform to within $10^{-6}$ generated in 514-bit blocks, with 147,456 seed bits, from the $H_{\text{min}}=0.030$, $\eta=0.537$ dataset. From the $H_{\text{min}}=0.030$, $\eta=0.535$ dataset we extracted a total of 94981 certified random bits uniform to within $10^{-6}$ generated in 514-bit blocks, using 147,456 seed bits.

\section{Discussion}
We demonstrate a steering-based one-sided device independent random number generator which produces certified random bits with the detection loophole closed. With heralding efficiencies for Alice above the steering threshold and below the threshold for device-independent methods, we perform randomness certification in an experimental regime where no randomness can be certified by DIQRNG. Another advantage of  the steering scheme is that the per-trial violation~\cite{superluminal} can be significant, even though the generation of entangled pairs from the source is random and has low probability ($\sim 0.1\%$) per pulse. This is because Bob is a trusted party, and thus a valid trial is defined whenever he receives a detection, regardless of the efficiency of his detection. This contrasts with DIQRNG with SPDC where, due to the large vacuum component in the two-mode state, the per-trial violation is low. We extract the random bits using a quantum-proof strong extractor---thereby demonstrating the first full implementation of a steering-based QRNG protocol. Although we close the detection loophole in this work, we assume no signaling between measurement devices and freedom-of-choice for measurement settings. By closing the locality loophole and strengthening freedom-of-choice in future implementations, these outstanding assumptions can be removed or weakened to increase the security of the randomness.   

Certified QRNGs with levels of device independence will  bring improved security to public randomness sources and private randomness for cryptographic applications. Here we have demonstrated a certified QRNG that can be extended to a strong-loophole-free one-sided device independent QRNG, and a viable randomness beacon. 
\section{Methods}

\begin{footnotesize}
\textbf{Steering Scenario.} We consider a scenario where, in each trial, a bipartite state $\rho_{\rm{AB}}$ is sent to two parties named Alice and Bob. Quantum steering can be framed as a task where Alice attempts to convince Bob they share a nonlocal state. Bob does not trust Alice and treats her measurement device as a black box with classical inputs $x$ and classical outcomes $a$, while Bob has full knowledge of his measurement device. 

Over many trials, Bob can perform tomographic measurements on his subsystem and Alice can report her settings and outcomes for measurements chosen randomly by Bob from the pre-agreed set. 

This allows Bob to construct a set of unnormalized quantum states that are conditioned on $a$ and $x$---this set is called an assemblage and consists of states:
\begin{equation}
\sigma_{a|x}=\rm{Tr}_{\rm{A}}\mathnormal{[(M_{a|x} \otimes \mathbb{1}_{B})\rho_{\rm{AB}}]},
\end{equation}
\noindent where $M_{a|x}$ is the measurement Alice performs on her subsystem. 

Alice can convince Bob of steering if the results of the experiment cannot be explained by any local hidden state (LHS) model where Bob receives states $\sigma_{\lambda}$, according to some local hidden variable $\lambda$ with deterministic probability distribution $D(a|x,\lambda)$. The existence of an LHS model gives assemblages a certain form:
\begin{equation}
\sigma_{a|x}=\sum_{\lambda} D(a|x,\lambda) \sigma_{\lambda}.
\end{equation}

Whether or not the assemblage admits an LHS model can be tested by implementing a semidefinite program (SDP)~\cite{sdpsteeringrev,steeringreview} to determine a quantity $\mu$:
\begin{equation}
\underset{ \{\sigma_{\lambda}\}}{\text{max}} \hspace{0.5cm}  \mu, \\
\end{equation}
$$ \text{s.t.}  \hspace{1cm}\sum_{\lambda} D(a|x,\lambda) \sigma_{\lambda}=\sigma_{a|x}   \hspace{1cm}   \forall{~a,x},\\$$
$$\sigma_{\lambda} \geq \mu\mathbb{1}  \hspace{1cm}    \forall{~\lambda}\\ .$$

\noindent 
This SDP finds the optimal set of local hidden states for Bob that are compatible with $\sigma_{a|x}$. For valid quantum states we must have $\sigma_{\lambda} \geq 0~~ \forall{~\lambda}$. 
Thus $\mu<0$ implies nonphysical states are needed and there can be no LHS model. 

This can be represented as a more traditional steering test by considering the dual minimization problem: 
\begin{equation}
\underset{ \{F_{a|x}\}}{\text{min}} \hspace{0.5cm}  \rm{Tr} \mathnormal{\sum_{a,x} F_{a|x}\sigma_{a|x}}, \\ 
\end{equation}
$$ \text{s.t.} \hspace{1cm} \sum_{a,x} F_{a|x} D(a|x,\lambda)\geq 0   \hspace{1cm}   \forall{~ \lambda},\\$$
 $$ \rm{Tr} \mathnormal{\sum_{a,x,\lambda} F_{a|x} D(a|x,\lambda)=} 1. \\$$ 
\noindent The dual variable, $F_{a|x}$, is a set of Hermitian operators called a steering functional, and defines a steering inequality:
\begin{equation}
\label{ineq} 
\beta := \rm{Tr} \mathnormal{\sum_{a,x} F_{a|x}\sigma_{a|x} \geq \beta _{LHS}=} 0,\\ 
\end{equation}
\noindent which is satisfied by LHS assemblages and violated by steering assemblages. We can see that the bound is $\beta _{LHS}=0$ by considering the first constraint in the dual problem and noting that, for an LHS model, $\sigma_{a|x} = \sum_{\lambda}  D(a|x,\lambda)\sigma_{\lambda}$. \\

\textbf{Randomness Extraction.} Trevisan’s extractor is a procedure which combines weak design and one-bit extractor algorithms~\cite{tmps2012}. Weak design is an algorithm that divides the uniform seed into a set of smaller bit strings with some overlap $r$, each of which is used by the one-bit extractor to extract one random bit from the weakly random string, called an $(n,k)$-source as it contains $n$ bits with $k=H_{\rm{min}}n$ min-entropy. Here we use quantum-proof polynomial hashing~\cite{tmps2012} (also known as an RSH or Reed-Solomon-Hadamard)  code for the one-bit extractor~\cite{instrumental2020}.
Our extractor outputs a random string of 
\begin{equation}
\label{ext}
m=\left \lfloor \frac{H_{\text{min}}n-4\log_{2}{\frac{1}{\epsilon}-6}}{r} \right \rfloor
\end{equation}
\noindent bits~\cite{tmps2012}, where $n$ is the length of the $(n,k)$-source and $\epsilon$ is the uniformity error tolerance chosen by the user---the sequence is statistically close to uniform with an error bounded by $\epsilon$. Clearly, $m<H_{\text{min}}n$, so there is some entropy loss dependent on the chosen $\epsilon$ value and the weak design overlap $r$. Here we use block weak design with $r=1$ which contributes no entropy loss ~\cite{tmps2012}. 
 Based on the given inputs, the code calculates $m$ and the required seed length $d$, performs the extraction and generates the random bit string, as well as additional information including seed length and runtime.  For the protocol to pass,  more randomness must be generated than is lost in extraction, such that $m \geq 1$. 
Our choice of one-bit extractor and weak design is ideal for low entropy loss, particularly with low entropy sources~\cite{tmps2012}. If the priority is fast extraction speed and throughput a different choice could be made at the expense of higher entropy loss. Speed is also dependent on the computational machine. 

\end{footnotesize}

\textbf{Acknowledgements}\\This work was supported in part by ARC grant DP210101651 and in part by ARC grant CE170100012. D.J.J.  acknowledges  support  by  the  Australian  Government  Research Training Program (RTP). We thank Lynden K. Shalm and Howard Wiseman for helpful conversations. \\

\noindent\textbf{Author contributions}\\
\noindent S.S. and G.J.P. conceived the idea and supervised the project. D.J.J. constructed and carried out the experiment with help from S.S.. A.P. and Y.W. assisted in the data analysis. S.X., B.X., I.R.B.  and S.R. developed the high-efficiency SNSPDs. All authors discussed the results and contributed to the manuscript.\\

\noindent\textbf{Additional information}\\
\noindent  Correspondence and requests for materials should be addressed to G.J.P.\\

\noindent\textbf{Competing financial interests}\\
\noindent The authors declare no competing financial interests.

%

\end{document}